\documentclass[10pt, conference, compsocconf]{IEEEtran}
\usepackage{graphicx}
\usepackage{url}
\ifCLASSINFOpdf
\else
\fi
\begin{document}
%
\title{Nighttime Light, Superlinear Growth, and Economic Inequalities at the Country Level}


\author{\IEEEauthorblockN{Ore Koren}
\IEEEauthorblockA{Department of Political Science, Indiana University\\
Bloomington, USA\\
Email: okoren@iu.edu}
\and
\IEEEauthorblockN{Laura Mann}
\IEEEauthorblockA{Jet Propulsion Laboratory, California Institute of Technology\\
Pasadena, USA \\
lauraerinmann@gmail.com} 
{\footnotesize\sffamily\textcopyright2018 California Institute of Technology}}


%


\maketitle

\begin{abstract}
	\noindent
Research has highlighted relationships between size and scaled growth across a large variety of biological and social organisms, ranging from bacteria, through animals and plants, to cities an companies. Yet, heretofore, identifying a similar relationship at the country level has proven challenging. One reason is that, unlike the former, countries have predefined borders, which limit their ability to grow ``organically.'' This paper addresses this issue by identifying and validating an effective measure of organic growth at the country level: nighttime light emissions, which serve as a proxy of energy allocations where more productive activity takes place. This indicator is compared to population size to illustrate that while nighttime light emissions are associated with superlinear growth, population size at the country level is associated with sublinear growth. These relationships and their implications for economic inequalities are then explored using high-resolution geospatial datasets spanning the last three decades.
\end{abstract}


%
\IEEEpeerreviewmaketitle

\section{Introduction}
Research shows that cities experience increasing returns to scale with respect to their size, a phenomenon known as ``superlinear'' growth \cite{Bettencourtetal2007,BettencourtWest2010}. According to data on U.S. cities, once logged, population size is associated with a $\sim 1.15$ unit increase in different urbanization measures---e.g., GDP, crime rate, patent applications---for a one unit increase in population when both measures are logged. 

Interestingly, however, the same does not appear to be true once one moves from the city to the country level. One possible explanation is that whereas cities experience organic growth and are unbounded, state borders are arbitrarily drawn, and hence place systemic constraints efficiency at the country level. Yet, to argue that state borders render scaled growth unlikely at the country level might be a case of ``throwing the baby out with the bathwater.'' Consider the difference between East and West Germany prior to unification. East German borders were preliminary (and arbitrarily) drawn in the Yalta Conference in 1945, and were decided \textit{de facto} based on where Soviet troops stopped in their progress west at the end of World War II \cite{Scott1998}. Despite this arbitrary division, the economic, political, and intellectual difference between East and West Germany could not have been more pronounced---whereas a strikingly censored, economically underdeveloped state evolved in the east, the west bloomed into an economic powerhouse and a haven for democracy. Countries, in other words, vary markedly based on their internal institutions, and whether and how the latter facilitate economic growth \cite{AcemogluRobinson2012}.

Is there a way to test whether growth at the country level is superlinear, sublinear, or both? Studies of cities and growth \cite{Bettencourtetal2007,BettencourtWest2010} rely on population size to approximate city size. Yet, population size is not divided ``efficiently'' within the state. While cities---unbounded---can attract the ``most efficient'' individuals (or those who have more resources) and experience superlinear growth, the implications for the countryside might be negative: those who are ``left behind'' in the countryside might even experience decreasing returns for scale as resources drain to the cities. As population movements between different countries still entail high costs, even for residents of internal single market economies such the European Union, the distribution of population within states is unlikely to reflect efficient allocation of resources and its effects on growth---measured, e.g., in gross domestic product (GDP)---due to the relatively arbitrary delineations of state borders. 

While population or geographic area might not be an efficient measure of country size, there is one indicator that does (presumably) reflect efficient allocation of resources within a given state: nighttime light emissions. Access to electricity is closely linked to local development and capacity. Indeed, recent research relied on satellite images of nighttime light emissions to approximate local development and capacity \cite{ChenNordhaus2011,Hendersonetal2012,WeidmannSchutte2017}. Nighttime light emissions reflects patterns of political mobilization, revenue mobilization, economic development, and even national security, which require efficient allocation of electricity and electric infrastructure, and is almost always governed by the state or by the (economic) agents sanctioned by it \cite{KorenSarbahi2017}. 

It is important to emphasize that nighttime lights are not synonymic with urbanization, and that one can observe high illumination even in rural areas (e.g., where large factories operate), while some cities, especially in developing states, might have large number of population but relatively low emissions \cite{KorenSarbahi2017}. Rather, the present paper compares two measures of scaled growth at the state level---one which is based on development and one which is based on population size---and identifies relevant implications.

\section{Nighttime Light and Scalability}

Whereas population growth is often unplanned (with the possible exemption of some states, e.g., China and North Korea)
and hence does not reflect the most efficient allocations of state capacity---and considering that changes in geographic area and state borders are quite rare---electricity infrastructure is often expended as to facilitate human activity, creativity, and growth \cite{WeidmannSchutte2017}. Infrastructure in these areas is also more likely to be maintained \cite{Harbers2015}. From this perspective, energy breeds information: the level of electricity provision is a better reflection of state ``size'' with respect to growth and prosperity compared with population size.

Highly complex structures---be they cells, cities, or states---require self-optimization to achieve efficient growth with respect to available resources. Such self-optimization is sustained via phenomena such as hierarchical branching networks, which allow the organism to achieve higher efficiency via self-similar structures \cite{Westetal1997,Bettencourtetal2007}. From this perspective, a good indicator of country ``size'' with respect to scaling will exhibit clear self-similarity across different orders of magnitude. 

\begin{figure}[!t]
	\centering
	\includegraphics[width=3in]{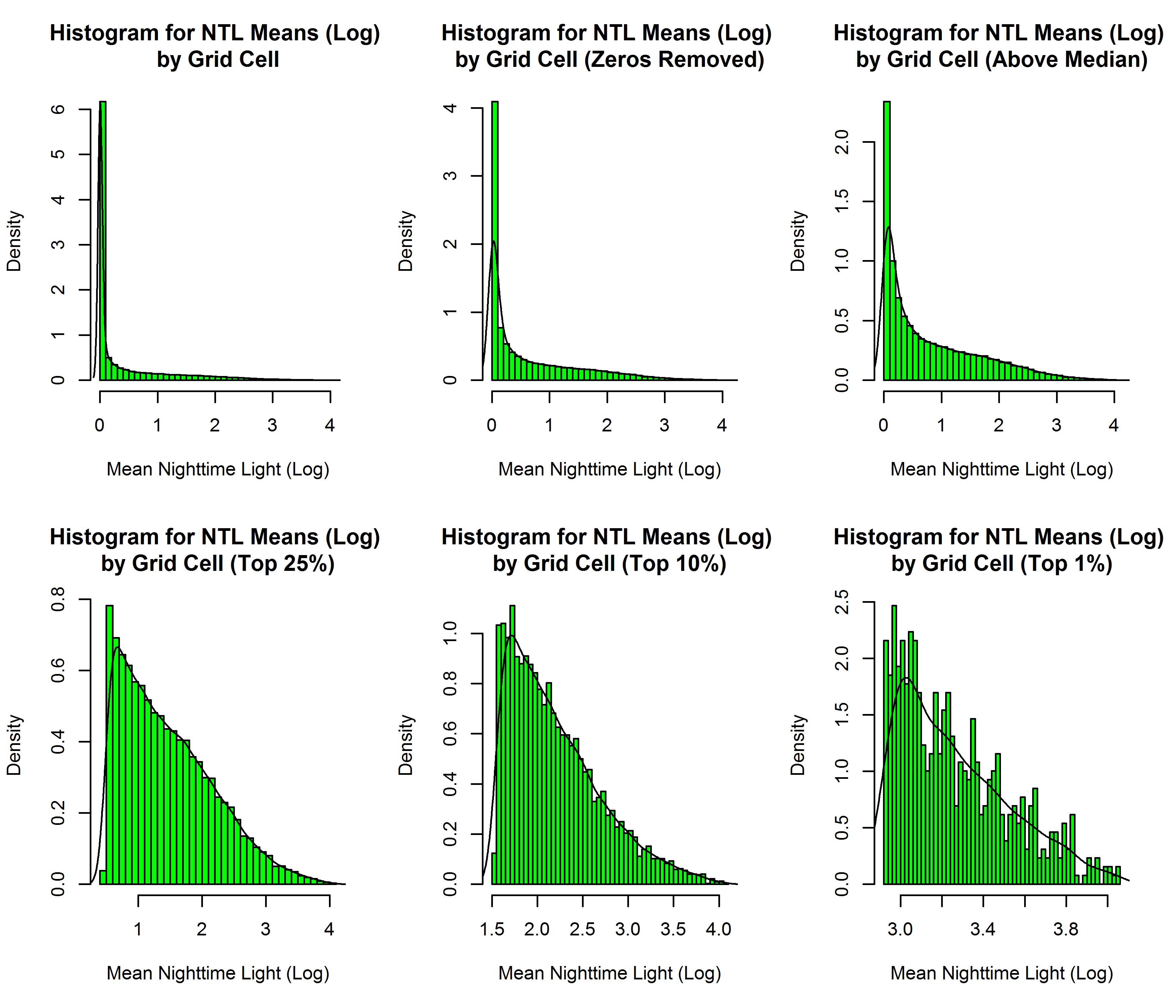}
	\caption{Histograms of Average NTL Frequencies by 0.5$\circ$ Grids (Logged), 1992-2013}
	\label{fig:logntl1}
\end{figure}

\begin{figure}[!t]
	\centering
	\includegraphics[width=3in]{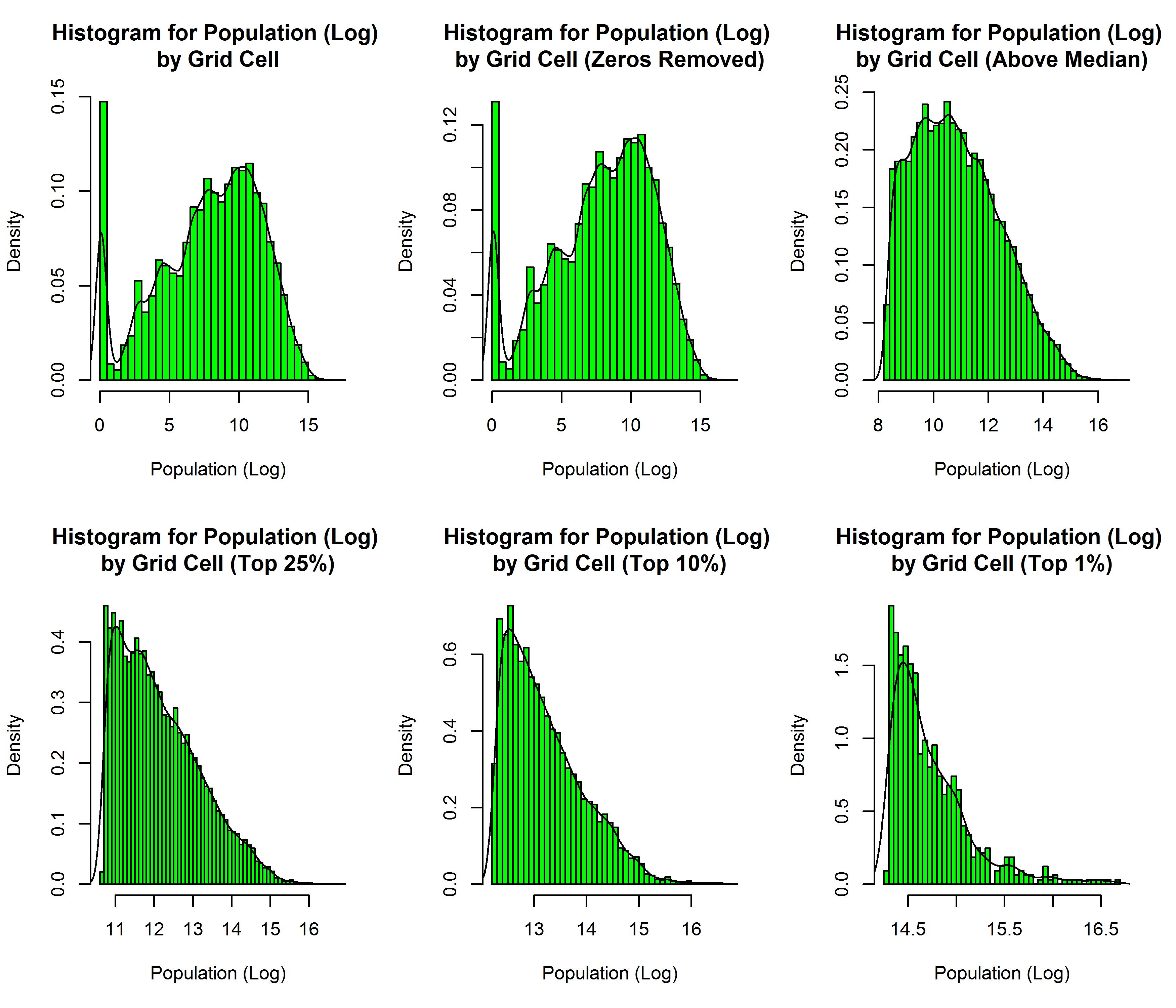}
	\caption{Histograms of Average Population Levels by 0.5$\circ$ Grids (Logged), 1995-2010}
	\label{fig:logpop1}
\end{figure}

Recent studies indeed find that nighttime light clusters show such scaling behavior, which is illustrative of networked systems \cite{Zhangetal2015}. To illustrate this last point, Figure \ref{fig:logntl1} plots frequency histograms for (logged) average nighttime light emissions, measured at the annual, high-granularity 0.5 degree ($\circ$) cell resolution\footnote{I.e., a ``square'' of approximately 55km x 55km at the equator, which increases in size as one moves toward the poles. The PRIO-Grid includes 64,818 such cells for any given year.} over the entire globe, which were averaged over the 1992-2012 period. Data on all cloud-free, composite nighttime light indicators used in this study were obtained from the PRIO-Grid dataset \cite{TollefsonForoStrandBuhaug2012}.
For comparison, Figure \ref{fig:logpop1} plots frequency histograms for (logged) average population levels, the standard size indicator in studies that focus on cities \cite{BettencourtWest2010}, measured at the same annual 0.5$\circ$ level over the 1995-2010 period and obtained from \cite{Nordhaus2006}. For each indicator, six plots are reported, showing values (i) across all grid cells, (ii) only non-zero values, (iii) cells with values above the median, (iv) cells with values in the top 75th percentile, (v) cells with values in the top 90th percentile, and (vi) cells with values in the top 99th percentile. For each of these figures, probability distributions are plotted to illustrate whether each measure maintains self-similarity across different thresholds.

Referring to Figure \ref{fig:logntl1}, the different plots illustrate a striking similarity of the probability density distributions of (logged) nighttime light levels across the different thresholds. The plot for (i) slightly resembles a power law distribution, but once zero values are removed, the nighttime light frequencies (ii-vi) exhibit remarkable similarity in their (linear) probability densities, regardless of the threshold used. This provides strong evidence to suggest that, indeed, nighttime light levels follow a power law distribution, and are hence scalable at higher levels of aggregation, such as the state.

In contrast, Figure \ref{fig:logpop1} illustrates that the probability density plots of the (log) population change shape across these different threshold values. The probability distributions shifts from a bivariate normal in (i) and (ii) to a skewed distribution in (iii), then to linear in (iv) and (v), and finally to a power-law like in (vi). This provides evidence to suggest that population size, while it might be a useful approximation of city size \cite{Bettencourtetal2007,BettencourtWest2010}, does not (log) linearly scale up to the country level, as illustrated in the ensuing section. 

\section{Testing Relationships at the Country Level}

The previous section illustrated that, tested against similar country-level measures of productivity/growth, nighttime light emissions and population size should yield divergent results. These expectations can be defined in mathematical terms. First, refer to the formula from \cite{Bettencourtetal2007}:

\begin{equation}
Y_t\ =\ Y_0N_t^{\beta}
\end{equation}

\noindent
Where $Y_t$ denotes material indicators of success at year $t$, $Y_0$ is a normalization constant, and $N_t$ is a measure of ``size,'' in the present case different population or nighttime light indicators. In this equation $\beta$ is the exponent: if $\beta>1$ (ideally, $\beta \simeq 1.15$) then it can be concluded that growth is superlinear (denoted for convenience as $\beta^{\uparrow}$); if $\beta<1$ (ideally, $\beta \simeq 0.75$) then it can be concluded that growth is sublinear (denoted for convenience as $\beta^{\downarrow}$). Taking the (natural) log of this equation and simplifying yields the following linear equation:

\begin{equation}
\ln(Y_t)\ =\ \ln(Y_0) + \beta\ln(N_t)
\end{equation}

This equation can be estimated using ordinary least squares (OLS), across different operationalizations of $Y_t$ and $N_t$. To operationalize material success, the present article relies on two distinct real gross domestic product (GDP) indicators, in a manner used in past research \cite{Bettencourtetal2007,BettencourtWest2010}. The first GDP indicator was obtained from the ``Expanded trade and GDP data'' \cite{Gleditsch2002}, which covers each year during the 1950-2011 period, and presumably provide a better approximation of GDP and population compared other, more widely used measures. The second indicator was obtained from the World Bank data \cite{WB2016}, which cover the 1960-2014 period. The corresponding population indicators used to evaluate the relationship between population size and GDP were also obtained from \cite{Gleditsch2002} and \cite{WB2016}, respectively.\footnote{Note that because the unit of analysis is the country-year, the plotted data in the correlations below appears as a large number of small diagonal clusters.} 

To evaluate how nighttime light emissions scale with respect to GDP, two different nighttime-light-based indicators are used. The first, reported above, measures \textit{average} annual nighttime light emissions within a given 0.5 degree grid cell, and is hence a measure of how much electricity/developed infrastructure was \textit{available} within a given country in year $t$ \cite{ChenNordhaus2011}. The second indicator measures the annual \textit{standard deviation} in emissions within a given 0.5 degree grid cell, and is hence a good approximation of how much new infrastructure was \textit{added} in a given country during year $t$. Both indicators were obtained from the PRIO-Grid dataset mentioned above \cite{TollefsonForoStrandBuhaug2012} and were measured at the same annual 0.5 degree cell level of resolution for the years 1992-2013, and both were aggregated to the country-year level to correspond to the GDP indicator's level of operationalization. 

\begin{figure}[!t]
		\centering
		\includegraphics[width=1\linewidth, clip=true, trim = 10mm 40mm 10mm 40mm]{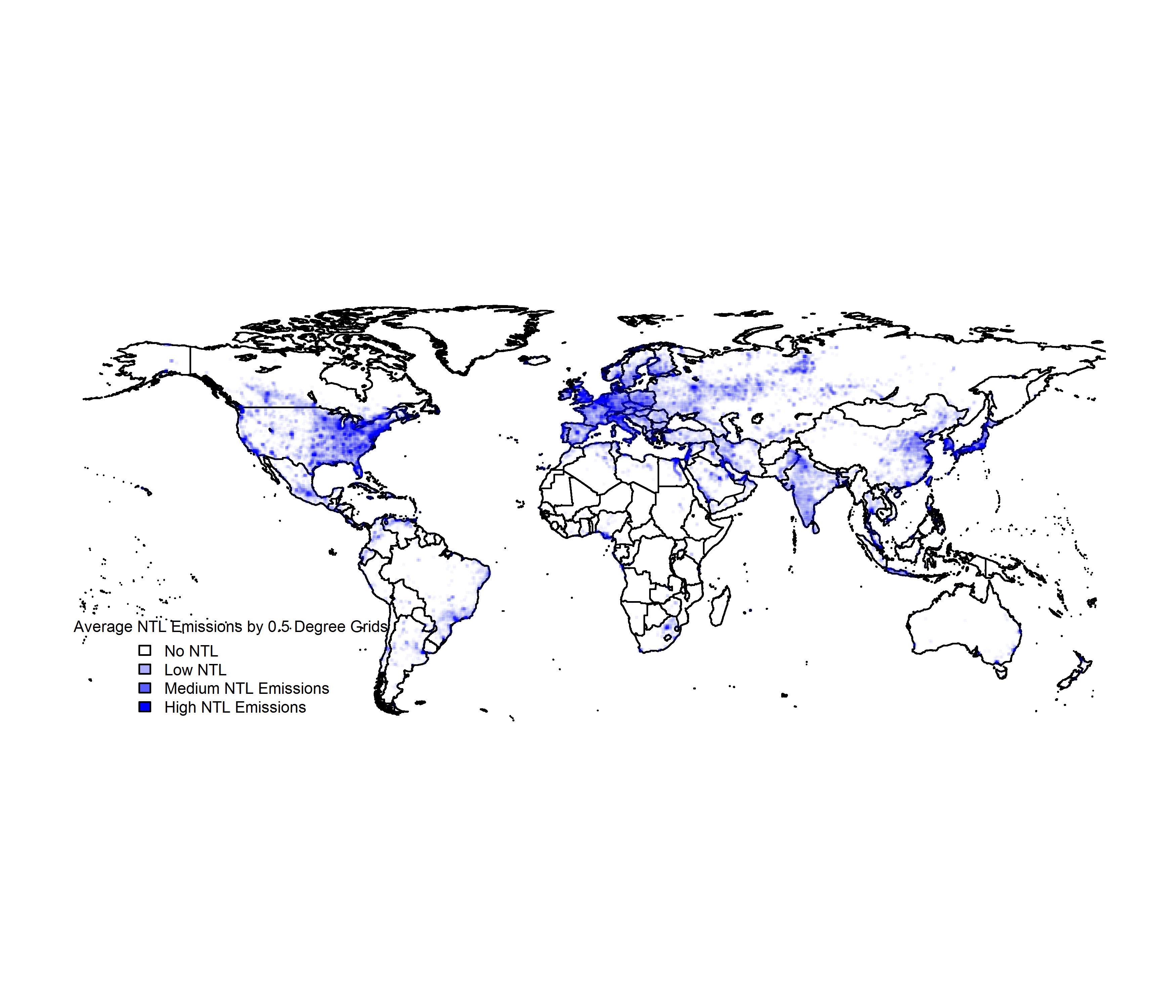}
		\caption{Average NTL Emissions by 0.5 Degree Grids, 1992--2013}
		\label{fig:ntlmapmean}
	\end{figure}

\begin{figure}[!t]
		\centering
		\includegraphics[width=1\linewidth, clip=true, trim = 10mm 40mm 10mm 40mm]{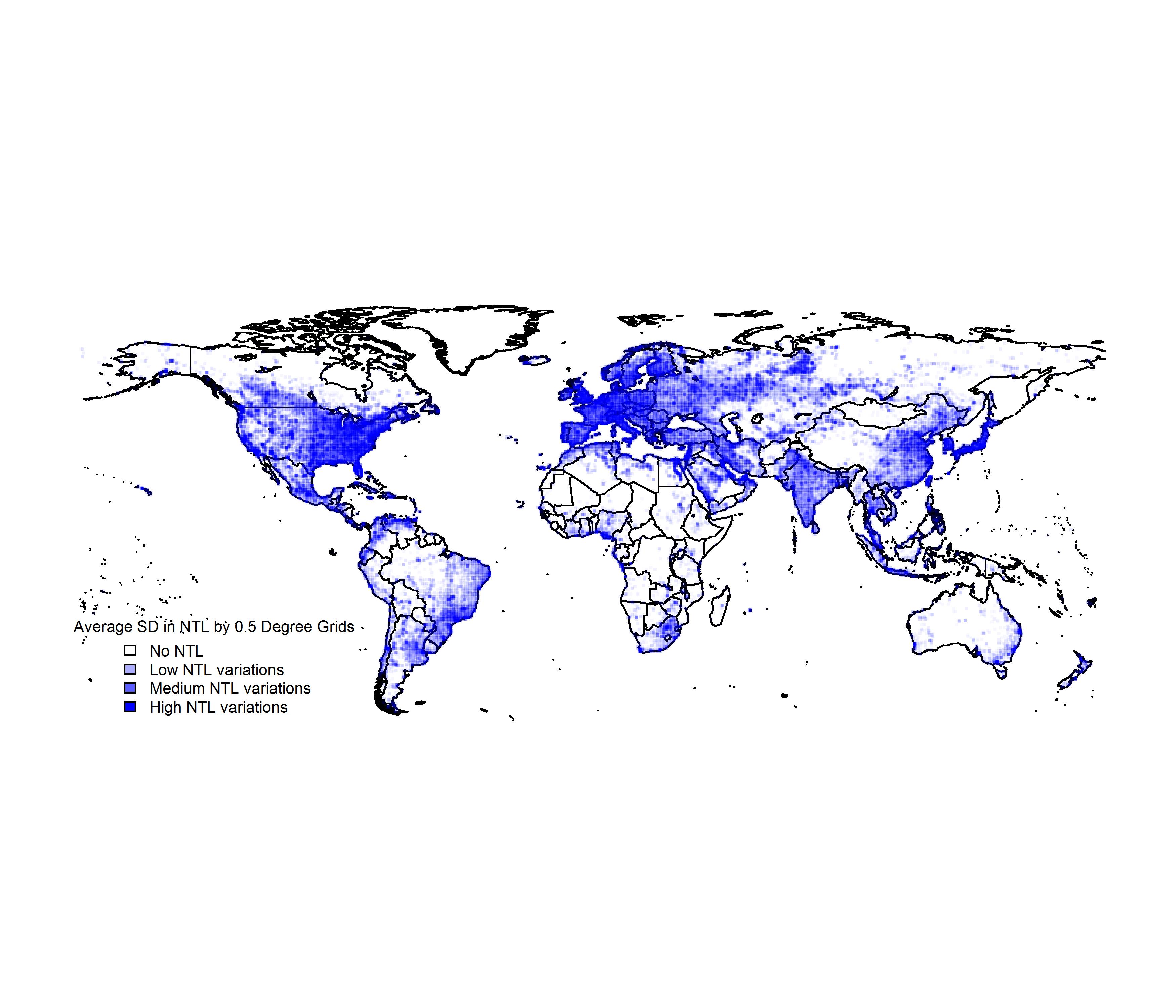}
		\caption{SD in NTL Emissions by 0.5 Degree Grids, 1992--2013}
		\label{fig:ntlmapsd}
	\end{figure}

\begin{figure}[!t]
		\centering
		\includegraphics[width=1\linewidth, clip=true, trim = 10mm 40mm 10mm 40mm]{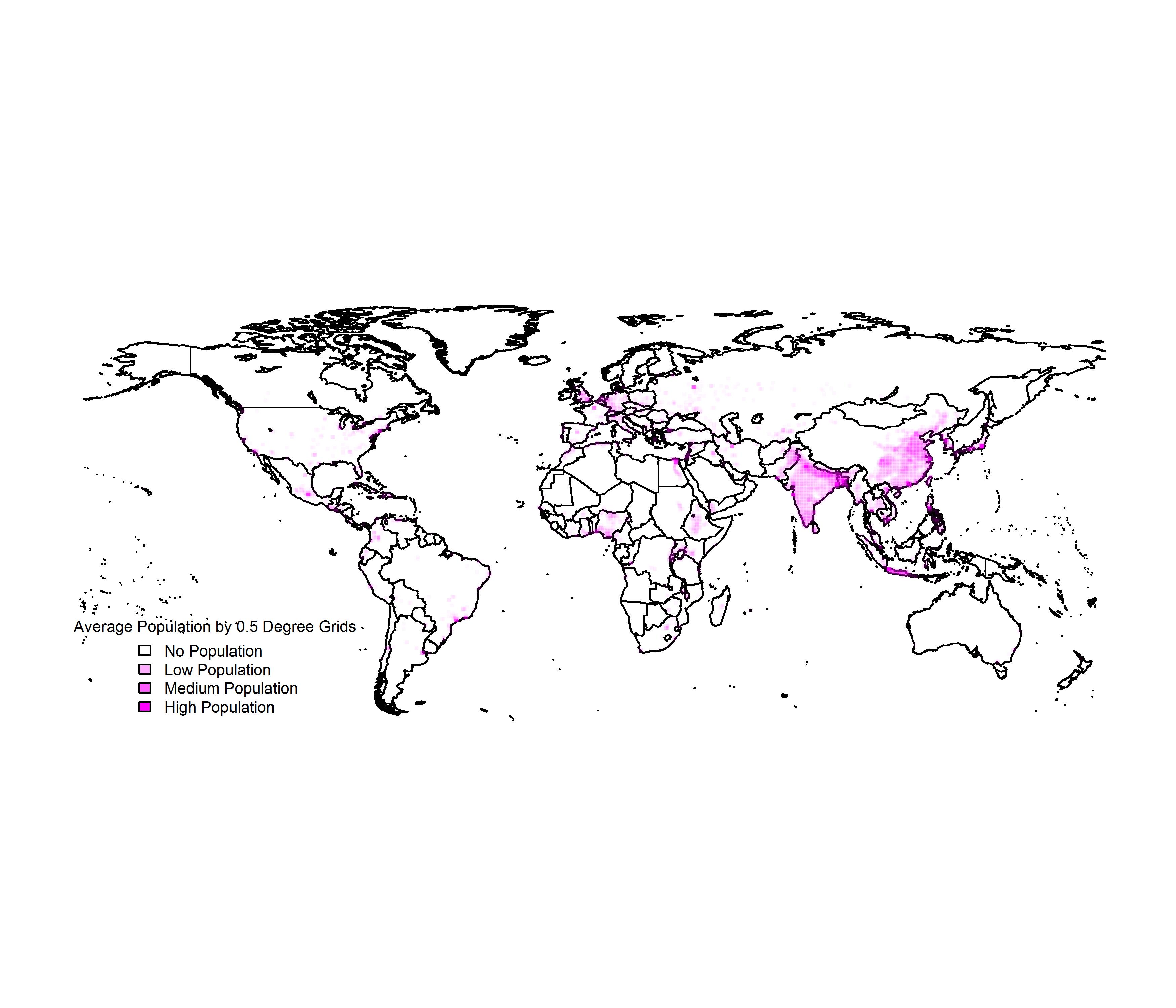}
		\caption{Average Population by 0.5 Degree Grids, 1995--2010}
		\label{fig:popmap}
	\end{figure}

For illustration, average and standard deviations (SDs) in nighttime light levels by 0.5 degree cell across the entire globe are plotted in Figures \ref{fig:ntlmapmean} and \ref{fig:ntlmapsd}, respectively, and were averaged for the 1992-2013 period. For comparison, population densities for each 0.5 degree cell (the same ones used in Figure \ref{fig:logpop1}) are plotted in Figure \ref{fig:popmap} and were averaged over the entire 1995-2010 period.\footnote{Considering the high population density in China and India, it might be hard to infer from this map where most people reside across the entire terrestrial globe. For a better illustration of the latter, a similar map where population size was logged is reported in the online appendix.}

\begin{figure}[!t]
	\centering
	\includegraphics[width=2.5in]{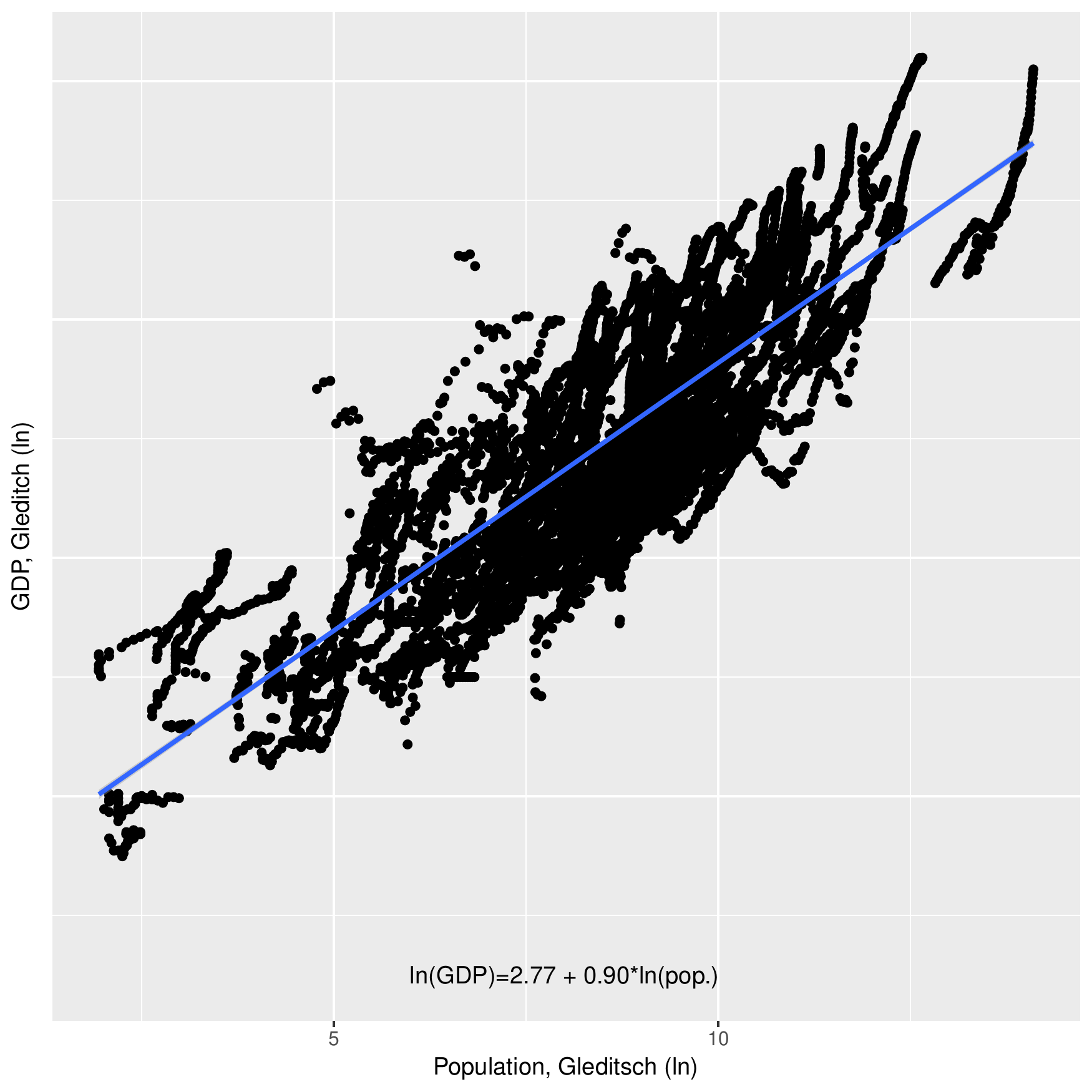}
	\caption{Global Regression Estimates, Population and GDP (Gleditsch 2002)}
	\label{fig:popcor1}
\end{figure}

\begin{figure}[!t]
	\centering
	\includegraphics[width=2.5in]{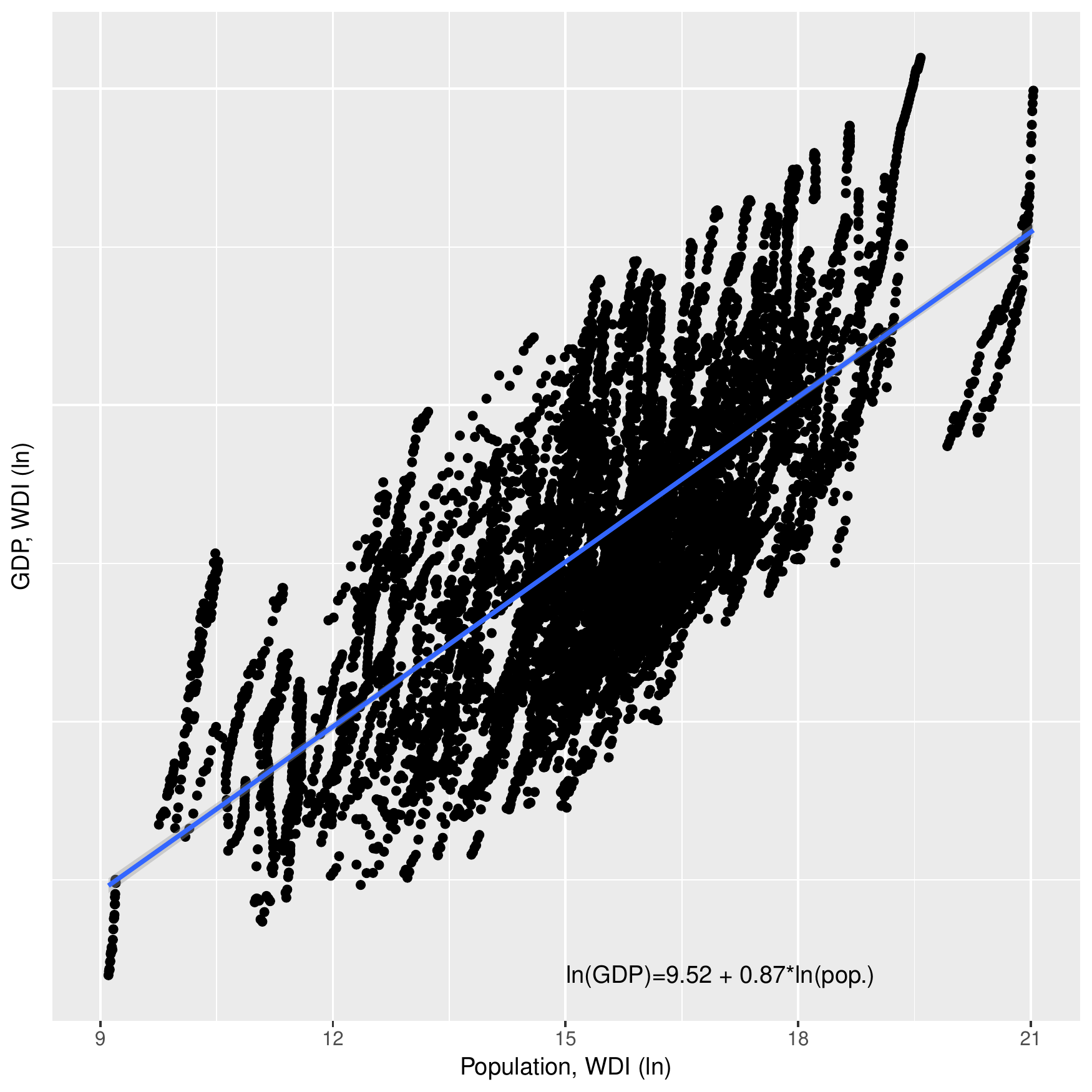}
	\caption{Global Regression Estimates, Population and GDP (World Bank 2016)}
	\label{fig:popcor2}
\end{figure}

\begin{figure}[!t]
	\centering
	\includegraphics[width=2.5in]{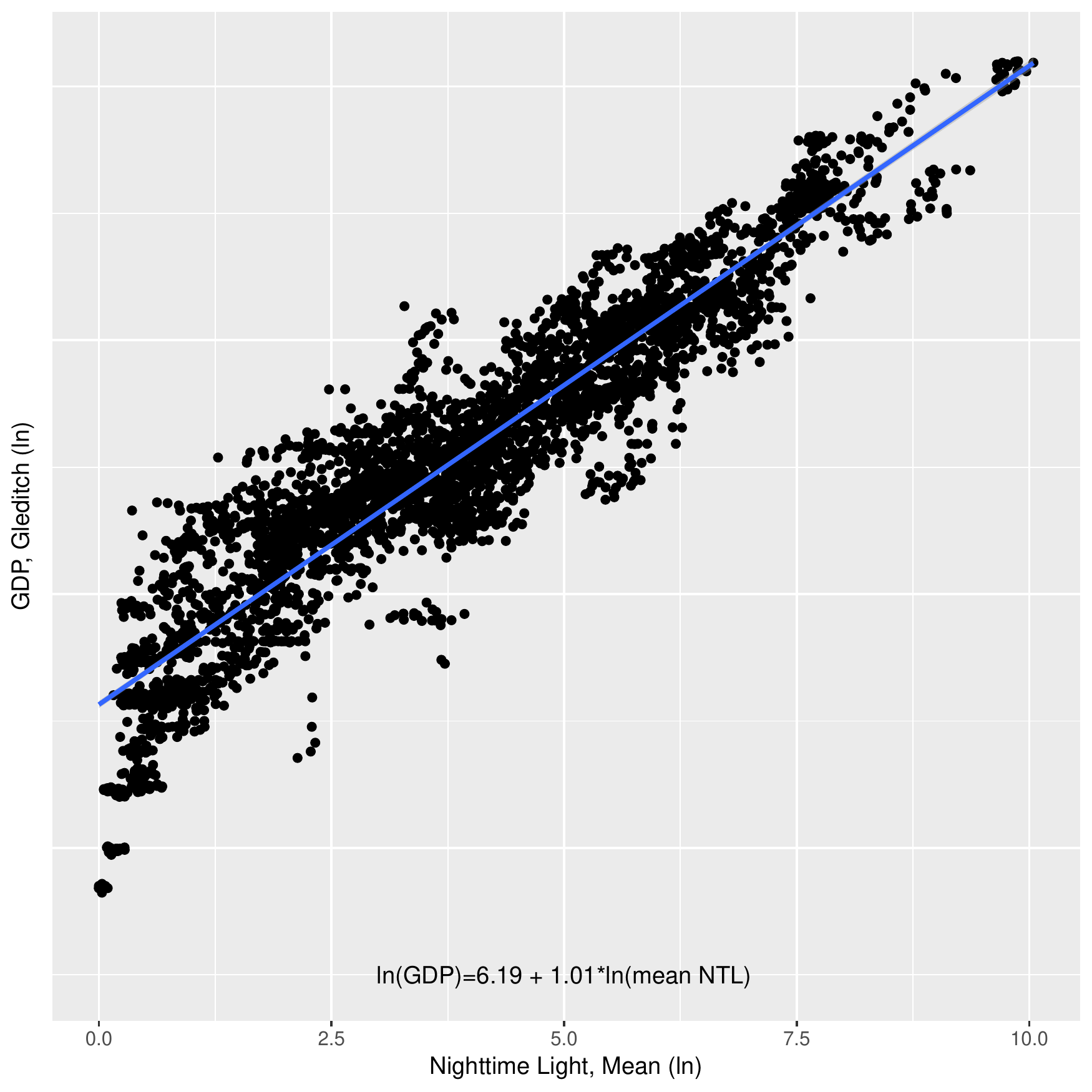}
	\caption{Global Regression Estimates, Nighttime Light Emissions (Mean) and GDP}
	\label{fig:nlcor1}
\end{figure}

The correlations between mean nighttime light and GDP, and nighttime light SDs and GDP, are plotted in Figures \ref{fig:popcor1}--\ref{fig:popcor2}, respectively. As both figures illustrate, population size scales sublinearly with respect to GDP, with (similarly sized) coefficients of $\beta^{\downarrow} \simeq 0.9$. Turning to the effects of nighttime light, Figure \ref{fig:nlcor1} illustrates that (natural log) mean nighttime light values have a $\beta^{\uparrow} \simeq 1.01$, while (natural log) standard deviation in nighttime light values have a $\beta^{\uparrow} \simeq 1.15$. Overall, the evidence strongly suggests that nighttime light emissions---as an approximation of local development and efficient energy allocations---are indeed a scalable measure of country ``size'' (in terms of energy availability) with respect to growth and economic prosperity.\footnote{Additional verification of this relationship using limited survey-based data on electricity provisions at the country level is provided in the online appendix.}

\begin{figure}[!t]
	\centering
	\includegraphics[width=2.5in]{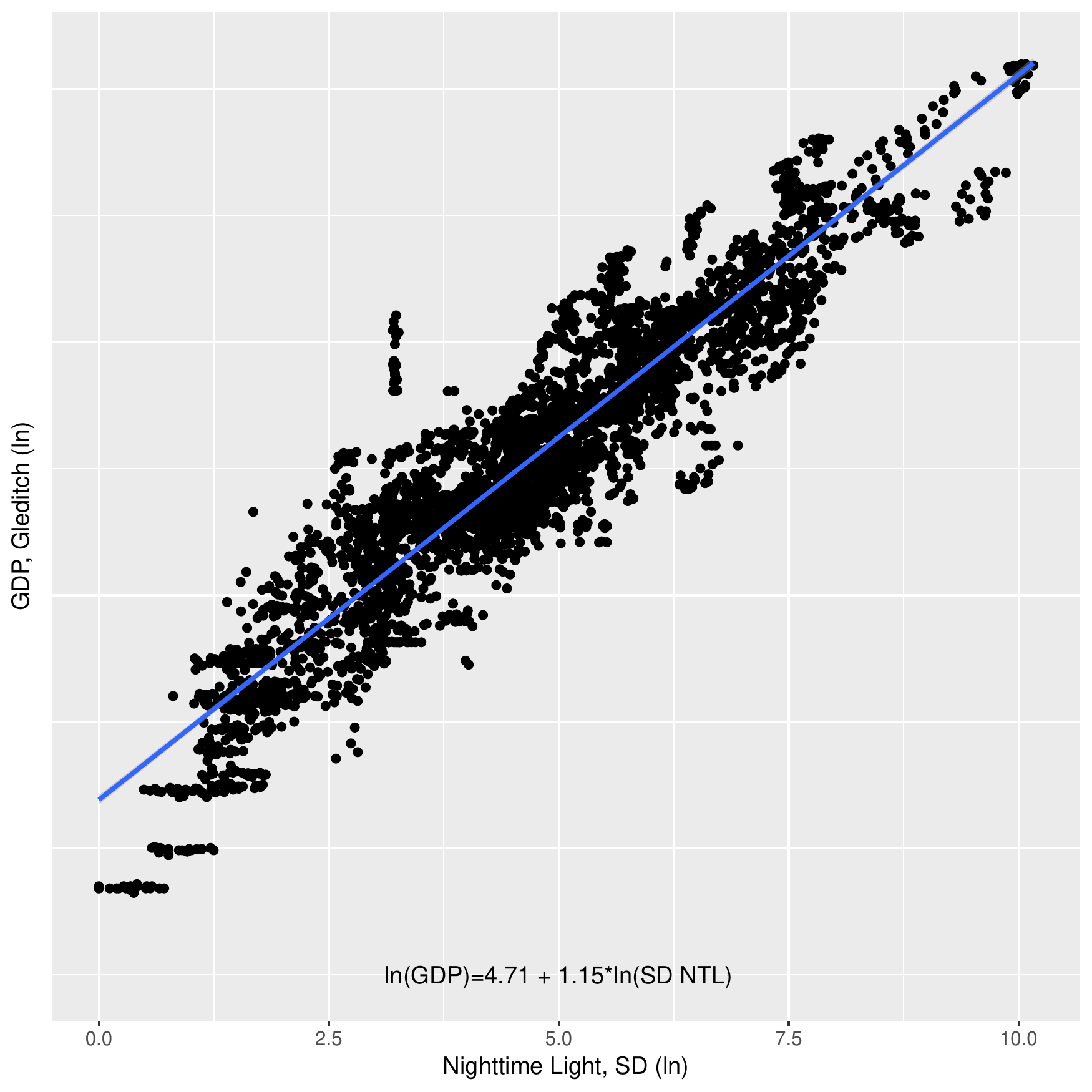}
	\caption{Global Regression Estimates, Nighttime Light Emissions (SD) and GDP}
	\label{fig:nlcor2}
\end{figure}

\section{Implications}

An implication of the superlinear scaling of nighttime light and the sublinear scaling of population at the country level is that such dynamics are likely to generate both ``winners'' and ``losers.'' Research suggests that high-efficiency urbanization leads to increased rates of economic development and knowledge creation, as resources ``drain'' toward exponentially growing urban centers \cite{Bettencourtetal2007,BettencourtWest2010}. The downside of these dynamics, the negative impact on peripheral urban areas and the rural countryside, can be daunting. This can also explain why nighttime light, as a measure of efficient energy allocation for the most successful activities, scales superlinearly at the country level, while population size, which is less mobile compared with energy allocations, scales sublinearly. As the preferences of governments and private actors favor highly-productive areas, and as those who have the capacity and the financial ability move to the center (cities), the periphery remains---on average---depleted, facing economic downturns, increasing inequality, and shortages in human capital.

\begin{center}
	\begin{figure}[!t]
		\centering
		\includegraphics[width=1\linewidth, clip=true, trim = 10mm 40mm 10mm 40mm]{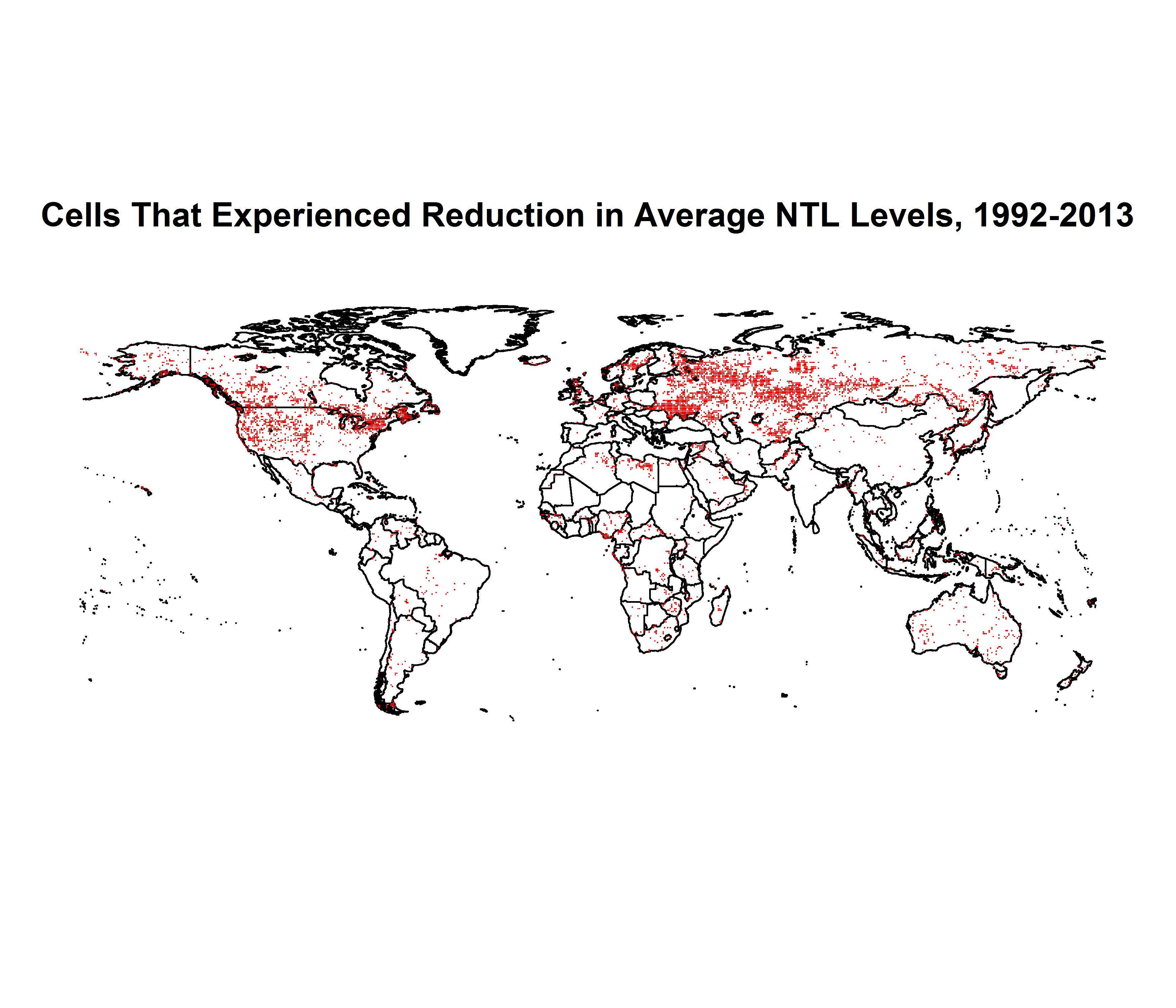}
		\caption{0.5 Degree Grid Cells that Experienced Reduction in Mean Nighttime Light between 1992 and 2013}
		\label{fig:ntlmapchange}
	\end{figure}
\end{center}

\begin{center}
	\begin{figure}[!t]
		\centering
		\includegraphics[width=1\linewidth, clip=true, trim = 10mm 40mm 10mm 40mm]{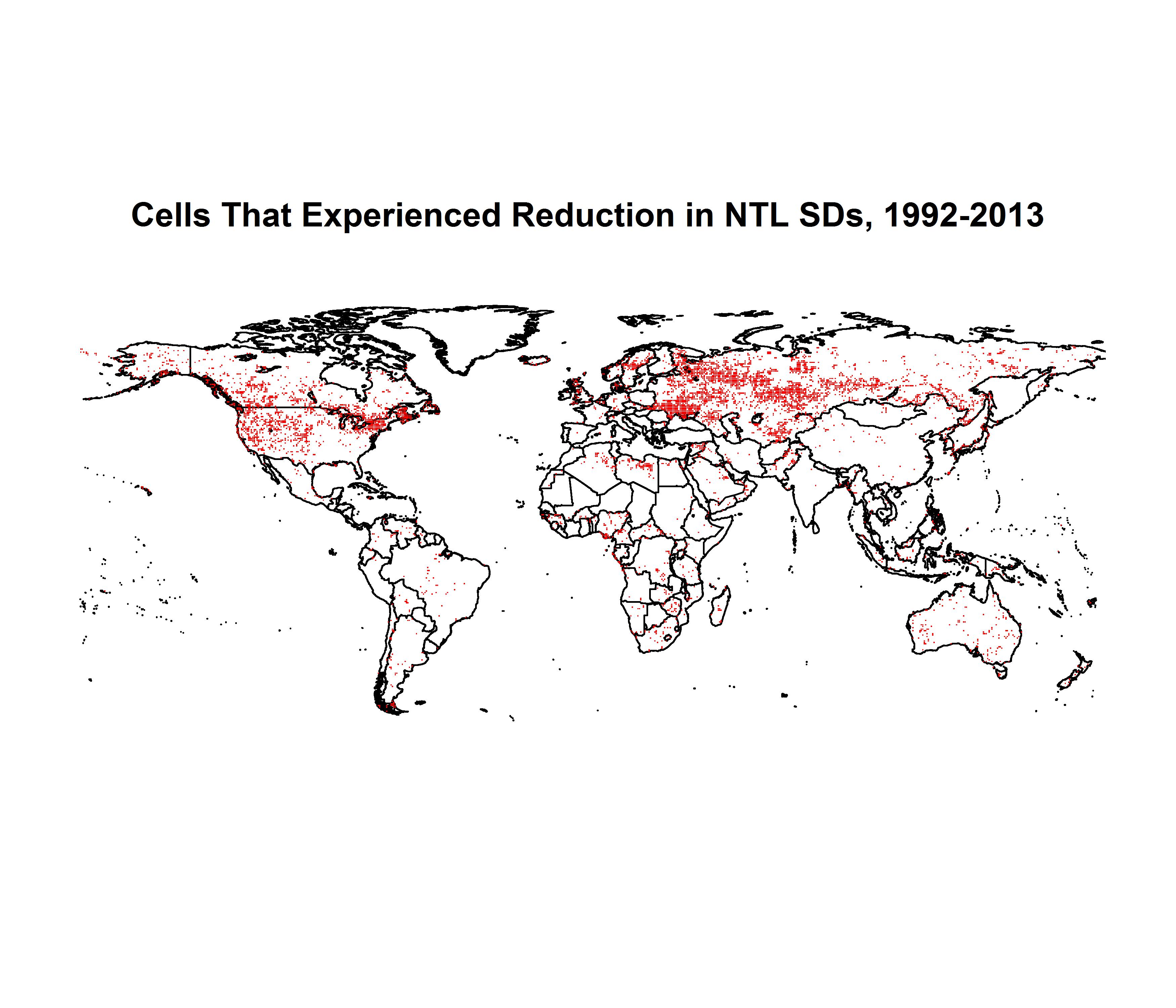}
		\caption{0.5 Degree Grid Cells that Experienced Reduction in Nighttime Light SDs between 1992 and 2013}
		\label{fig:ntlsdmapchange}
	\end{figure}
\end{center}

To evaluate whether changes in the global distribution of nighttime light over the last decades have had a negative impact on some regions but not others, Figure \ref{fig:ntlmapchange} plots all the 0.5$\circ$ cells that experienced a \textit{decrease} in average nighttime light levels between 1992 and 2013. This map strongly suggests that indeed, nighttime light levels decreased mostly in rural areas or regions facing economic decline. Overall, of the total of 63,415 grid cells on which information on nighttime light was available, 5,086 cells experienced a decrease in \textit{average} nighttime light emissions. As Figure \ref{fig:ntlmapchange} illustrates, the region primarily affected were states of the former Soviet Union, which experienced significant decline in economic output during the period of observation \cite{Brizgaetal2013}, as well as the American ``Rust Belt''  \cite{Alderetal2014} and industries east of the Pacific Coast and west of the Mississippi; the Rocky and maritime provinces of Canada; rural parts of sub-Saharan and northern Africa; and rural parts of Australia, among others. The results remain visually unchanged when negative changes in nighttime light emissions SDs are plotted in Figure \ref{fig:ntlsdmapchange}, although more cells experienced decreases in nighttime light variability, i.e., decreases in development activity (a total of 5,890). 

\begin{center}
	\begin{figure}[!t]
		\centering
		\includegraphics[width=1\linewidth, clip=true, trim = 10mm 40mm 10mm 40mm]{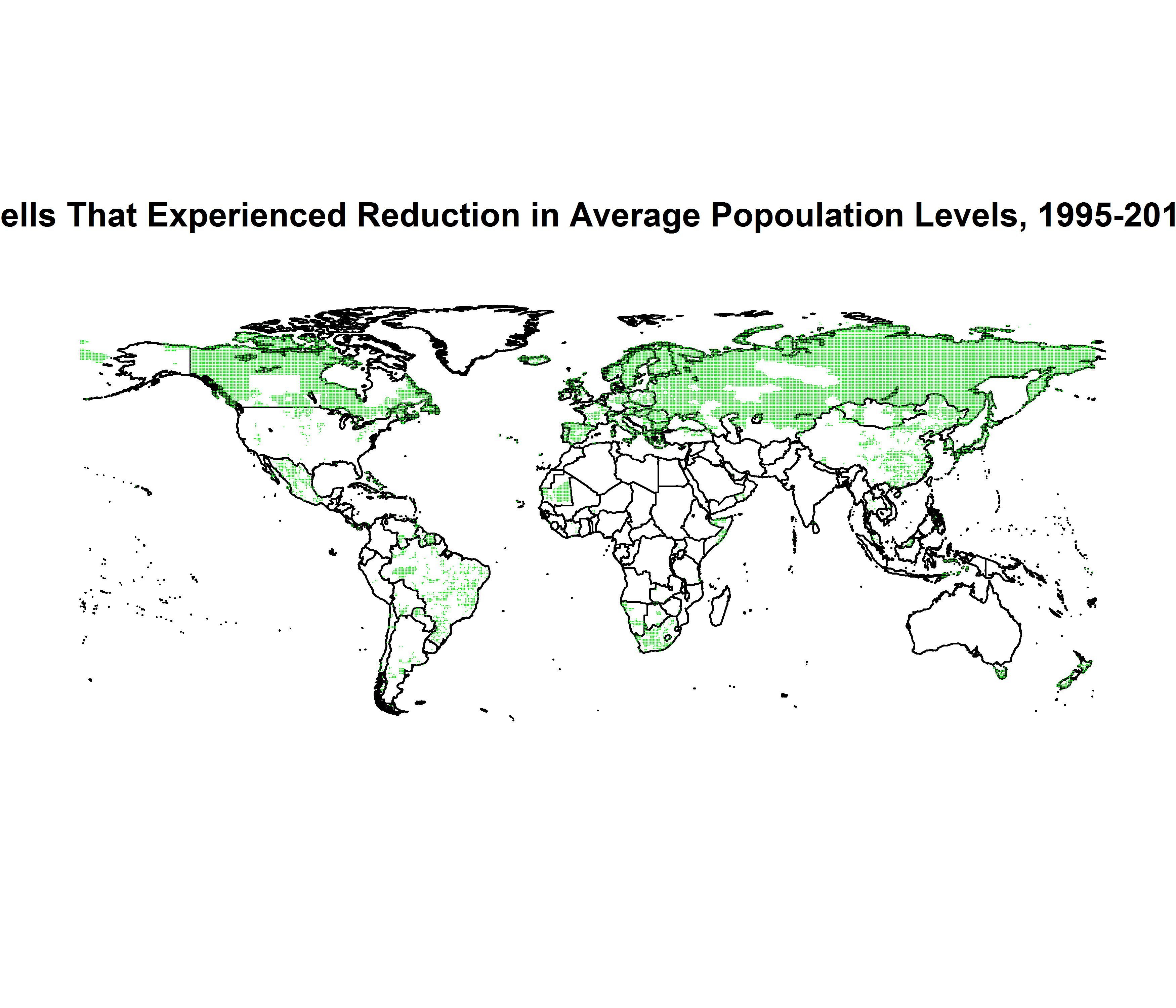}
		\caption{0.5 Degree Grid Cells that Experienced Reduction in Population Size between1995--2010}
		\label{fig:popmapchange}
	\end{figure}
\end{center}

Interestingly, Figure \ref{fig:popmapchange}, which plots all the 0.5$\circ$ grid cells that experienced decreases in population size between 1995 and 2010, illustrates that while some overlap exists, many of these local decreases in nighttime light were not the result of decreases in population. Overall, of the 64,818 grid cells in the PRIO-Grid, 24,067 experienced population decreases between 1995 and 2010 (while the total population for all grid cells increased by 1.16 billion people over the same period), the vast majority of which located in Canada and formerly Soviet states. Note that due to the way this population variable was constructed, local changes in population were broadly extracted from general country trends, excluding regions where information on such changes was not available,\footnote{E.g., in the northern part of the Canadian provinces Alberta and Saskatchewan.} which might obscure some the local level relationships between nighttime light and population. Nevertheless, while overlap between decreases in nighttime light emissions and decreases in population exist in some regions, most noticeably in the former Soviet Union, other regions exhibit marked divergence. 

\begin{center}
	\begin{figure}[!t]
			\centering
			\includegraphics[width=3in]{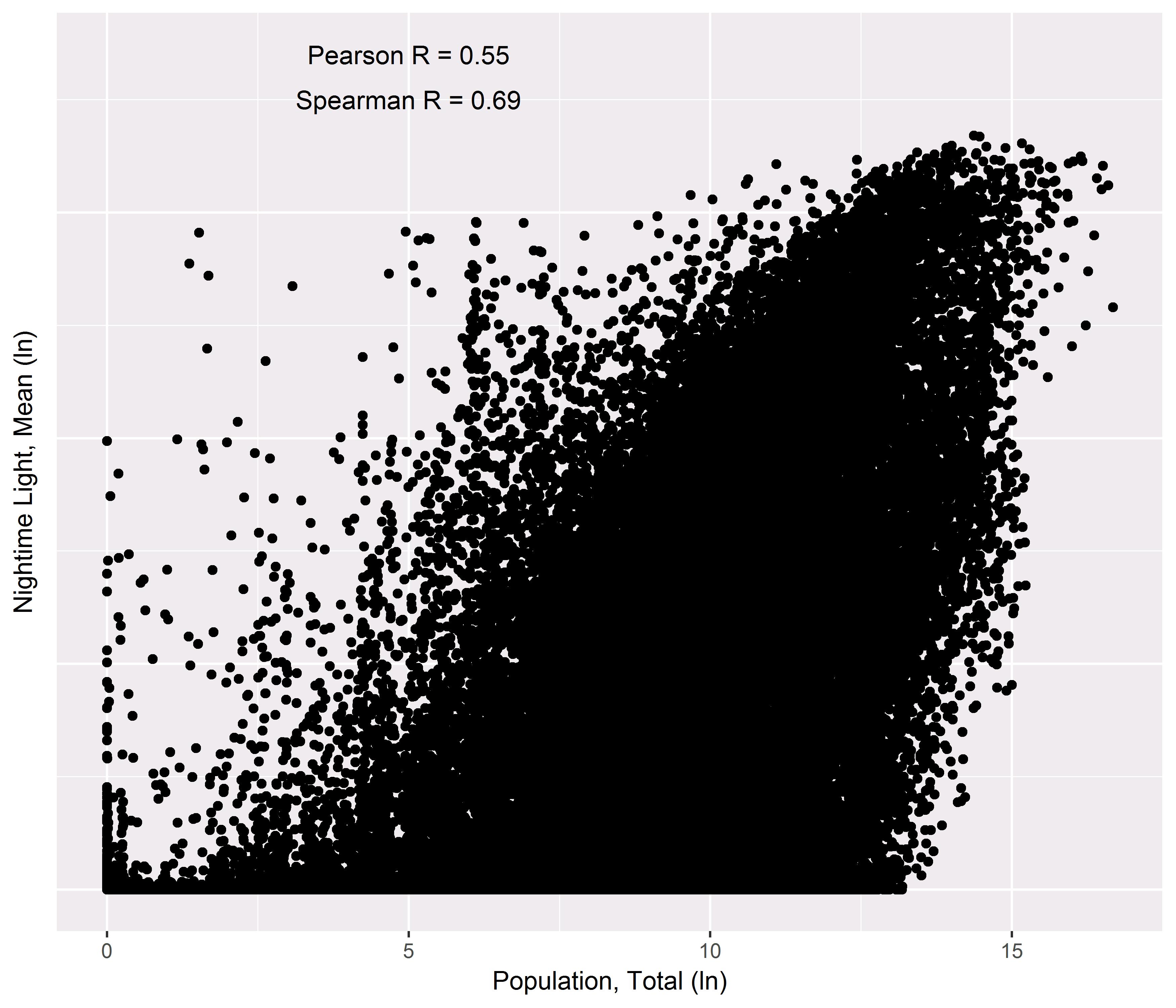}
		\caption{Correlation between Population Size and Mean Nighttime Light Levels, Average for 1992--2013}
		\label{fig:nlpopcor1}
	\end{figure}
\end{center}

\begin{center}
	\begin{figure}[!t]
		\centering
		\includegraphics[width=3in]{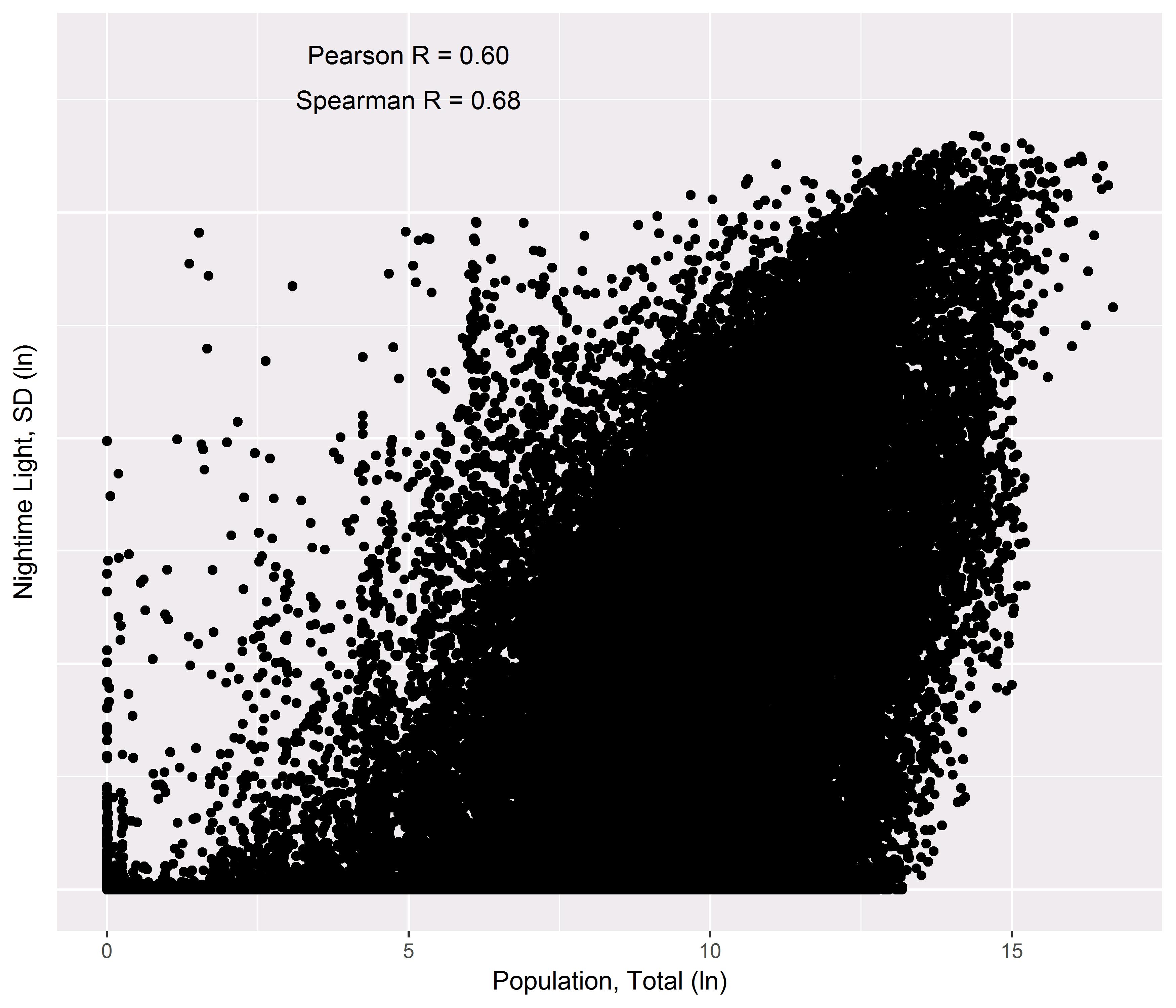}
		\caption{Correlation between Population Size and Nighttime Light SDs, Average for 1992--2013}
		\label{fig:nlpopcor2}
	\end{figure}
\end{center}

The American ``Rust Belt'' for instance, especially the area covering upstate New York, Pennsylvania, and Ohio, shows large decreases in average nighttime light emissions and their SDs, while experiencing only mild, highly-localized decreases in population size. From a complementary perspective, both western Europe and eastern China experienced population decreases, but show no declines in nighttime light emissions, as explained by the fact that these regions went through urbanization and development over the same period \cite{Shaker2015,Cimolietal2009}. These discrepancies are also illustrated by Figures \ref{fig:nlpopcor1}-\ref{fig:nlpopcor2}, which show that a large number of high-population areas are located in cells that rank relatively low on the nighttime light emissions scale (operationalized both as mean levels and SDs). 
As resources, qualified individuals, and industries move toward more productive areas, primarily coastal cities, within-country inequalities deepen.

\section{Discussion}

The enormous complexity of states combined with the tendency of their borders to be ``inorganic,'' that is defined by factors external to their organic growth and development, has prevented past research from identifying scaled growth in these entities. Hence, while scaled growth has been shown across a wide range of phenomena, ranging from cells and bacteria, through animals and plants, to cities and companies \cite{West2017}, countries have been considered a possible exception. By focusing on nighttime light---a proxy of infrastructure and development---as a measure of organic growth, this paper offered a way of extending the theory of growth from bacteria through cities, all the way to states, a remarkable achievement for complexity studies. 

Nighttime light emissions are used to facilitate the generation of information, and hence follow an organic path of allocations, with the most productive countries showing the highest levels of nighttime light emissions. As a result, across the entire globe both average nighttime light levels and their variability were shown to exhibit superlinear growth at the country level. As was shown in Figure \ref{fig:logntl1}, nighttime light emissions also exhibit remarkable self-similarity in their global distribution across different thresholds, indicating a universality of human activity despite enormous variability in national form. In contrast, population, the most widely used measure of ``size'' in past studies \cite{BettencourtWest2010,West2017}, was shown here to be sublinearly associated with GDP. 

The logic underlying the use of nighttime light as a measure of organic growth as advocated here building on \cite{Bettencourtetal2007}, who argue that ``increasing rates of innovation, wealth creation...suggest flows of these quantities from places where they are created faster (sources) to those where they are produced more slowly (sinks) along an urban hierarchy of cities dictated, on average, by population size.'' One potential objection is that such growth is still, ultimately, constrained by the state's physical area. While it might be true that there is a point at which countries might run out of physical space for further growth, within-country spaces can be reallocated as to facilitate more economic activity, and hence generate more nighttime light emissions. Singapore, Hong Kong, and Taiwan are some examples of such states, but the same is also true for larger countries such as the United States, where---as Figure \ref{fig:ntlmapchange}--\ref{fig:ntlsdmapchange} illustrated---development and electricity allocation shifted to favor areas where more creative activity takes place. 

The practical implications of these findings highlight the importance of measuring and understanding the impact of superlinear growth on areas whose resources dwindle in the process, and where people cannot move to areas where more economic activity takes place---especially large cities, where the cost of living is often high---and hence fail to realize their potential. If we assume that scaled growth follows the laws of nature, we might also be forced to accept that allocating resources to more efficient venues via a process of optimization can result with the decay and even demise of ``subefficient'' parts. States might experience superlinear growth, while at the same time large segments of their respective populations, usually in the lower income tiers, will grow sublinearly, as was illustrated in Figures\ref{fig:popcor1}--\ref{fig:popcor2} and \ref{fig:nlpopcor1}--\ref{fig:nlpopcor2}. This suggests that economic and social optimization are not inherently good.

This paper's findings suggest future directions to be explored in generalizing the empirical observations made here to other quantities, and to our understanding of scaling and inequality. One such venue relates to the environmental impact of development and infrastructures, and how reallocating infrastructure and production from some regions to others can generate negative (and positive) externalities. A second direction involves investigating how the social dynamics identified here create two groups: those who benefit from access to more infrastructure, and those who lose from it. This knowledge will suggest paths along which a future where economic activity and innovation can lead to improvements in human living standards across both the center and the periphery.

If true, these findings can also help to explain political polarization across developed and underdeveloped areas within a given country \cite{ScalaJohnson2017}. This in turn can open new lines of inquiry into the causes of inequality, and help policymakers working to ameliorate these issues and ensure that more individuals get to enjoy the benefits of superlinear growth rather than suffer its sublinear implications.

\section{Materials and Methods}

The datasets used for country level analysis where obtained from the World Bank and from the ``Expanded trade and GDP data'' and discussed above. All data at the annual 0.5$\circ$ level were obtained from the PRIO-Grid, and also discussed above. Fits to country-level data in Figures \ref{fig:nlcor1}--\ref{fig:popcor2} were performed by using ordinary least-squares with a correction for heteroskedasticity at the country level using the R software package. Results remained practically unchanged when no corrections for heteroskedasticity were implemented. The online appendix is available at: \url{http://www.orekoren.com/wp-content/uploads/2018/07/Online-Appendix-Scaled-Paper-06_28_18.pdf}.

\section{Acknowledgments}
The research was carried out at the Santa Fe Institute , for Laura Mann under a training program funded by Caltech.

\end{document}